\documentclass[conference]{IEEEtran}
\IEEEoverridecommandlockouts
\usepackage{amsmath,amsfonts}
\usepackage{algorithmic}
\usepackage{array}
\usepackage[caption=false,font=normalsize,labelfont=sf,textfont=sf]{subfig}
\usepackage{textcomp}
\usepackage{stfloats}
\usepackage{url}
\usepackage{verbatim}
\usepackage{graphicx}
\usepackage[normalem]{ulem}
\usepackage{amssymb}
\usepackage{listings}
\usepackage{multirow}
\usepackage{algorithm}
\usepackage{makecell}
\usepackage{booktabs}
\usepackage{color}
\def\BibTeX{{\rm B\kern-.05em{\sc i\kern-.025em b}\kern-.08em
    T\kern-.1667em\lower.7ex\hbox{E}\kern-.125emX}}
\begin{document}

\title{Synthesizing Proxy Applications\\ for MPI Programs}

\author{
\IEEEauthorblockN{
    Jiyu Luo,
    Tao Yan, 
    Qingguo Xu, 
    Jingwei Sun$^{*}$\thanks{* Corresponding author}, 
    Guangzhong Sun
}
\IEEEauthorblockA{\textit{University of Science and Technology of China, China}}
\IEEEauthorblockA{\{luojiyu, yantao98, xuqg\}@mail.ustc.edu.cn, \{sunjw, gzsun\}@ustc.edu.cn}
}

\maketitle

\begin{abstract}
Proxy applications (proxy-apps) are basic tools for evaluating the performance of specific workloads on high-performance computing (HPC) systems.
Since the development of high-fidelity proxy-apps, which exhibit similar performance characteristics as corresponding production applications, is labor-intensive, synthetic proxy-apps are created as a useful supplement to manually developed proxy-apps.
To thoroughly resemble performance characteristics of HPC applications represented by Message Passing Interface (MPI) programs, we propose Siesta, a novel framework to automatically synthesize proxy-apps based on communication-computation traces.
Given an MPI program, Siesta synthesizes parameterized code snippets to mimic computation behaviors in different execution periods, and combines the code snippets and MPI function records into an event trace.   
It then extracts program behavior patterns from the trace as grammars and finally transforms the grammars into a synthetic proxy-app.
We evaluate the proposed methods on representative MPI programs with various environments.
The results show that our synthetic proxy-apps can precisely approximate the performance characteristics of MPI programs.
\end{abstract}


\maketitle

\section{Introduction}
Proxy applications (proxy-apps) are basic tools for evaluating high-performance computing (HPC) systems.
They enable convenient system design space exploration without complicated library dependency \cite{date17minime,taco18SynchroTrace,tpds21sample}. They are also used for performance modeling and estimating \cite{tpds23pilgrim,jpdc22Var,tc16zhang},
and can play a role as benchmarks representing proprietary workloads \cite{pact17lizy,cgo17bench,jpdc19zhang}.
However, developing HPC proxy-apps is costly \cite{tc15minime,ijhpca18miniapps}. To precisely portray the complicated interaction between HPC systems and applications, the development requires intensive efforts from both application domain (e.g., physics, chemistry) experts and HPC developers. As a result, proxy-apps have difficulty in following the evolution of real-world HPC applications \cite{sigmetrics15apprime}.

Synthetic proxy-apps are constructed to approximate the performance characteristics of given programs, without performing any useful computation.
A typical process of generating synthetic proxy-apps mainly consists of two phases: recording program behaviors, and replaying the record.
Related recording and replaying techniques have been originally proposed for deterministic debugging purposes \cite{sfi2018survey}, but recent studies show that they are also effective for synthesizing proxy-apps \cite{iiswc14kim, tc16Zhai}.
Synthetic proxy-apps can leverage detailed performance profiling data and achieve promising performance coverage in many cases \cite{pact17lizy, sosp21workload}.
Since synthetic proxy-apps are generated from automated processes, they can be updated frequently and keep up closely with extensive real-world applications.
Besides, compared with general-purpose benchmarks, like SPEC CPU \cite{spec_cpu}, NPB \cite{ijhpca91bailey}, HPCG \cite{hpcg}, synthetic proxy-apps can mimic any given applications, satisfying customized demand.

The major challenge in synthesizing proxy-apps is extracting HPC applications' behavior patterns from a large amount of trace data.
For instance, tracing a single execution of LULESH \cite{IPDPS13LULESH} on less than 1,000 processors can produce hundreds of gigabytes of trace data \cite{tpds22sun}.
Existing methods mainly focus on communication trace \cite{sigmetrics15apprime,tc16zhang,jpdc19zhang,tpds21schulz,sc21pilgrim,tpds22sun}, since communication events constitute the skeleton of an HPC application. 
Meanwhile, computation behaviors are often overlooked.
Computation behaviors are more diverse and difficult to trace. 
As a compromise, they are usually recorded as time intervals and replayed by calling sleep functions or empty loops \cite{sigmetrics15apprime,tc16zhang,jpdc19zhang,tpds21schulz,tpds22sun}. 
Besides, lossy compression is often used to further reduce the size of communication information, of which the original communication parameters and timing information are approximated by histogram \cite{ipdps12frank,ipdps17frank, sc21pilgrim, tpds23pilgrim}, clustering \cite{jpdc17frank,ipdps18chameleon}, and stochastic process \cite{sigmetrics15apprime,tpds17astro,tpds22sun}.
Due to the ignorance of computation details and the lossy representation of communication trace, current synthetic proxy-apps are difficult to comprehensively resemble the performance characteristics of HPC applications.

To address these issues, we propose Siesta, a novel framework to automatically synthesize proxy-apps for HPC applications.
Since Message Passing Interface (MPI) is the leading programming standard for HPC, we focus on MPI programs in this study.
In general, given an MPI program without source code, we trace the program's communication and computation events and then convert the trace to a certain programming language (e.g., C language) code.
The generated code does not output meaningful results, but it mimics the communication and computation of the given MPI program and exhibits similar performance characteristics.

By extracting nested loops and SPMD (single program multiple data) patterns in the trace using a grammar-based compression method \cite{nevill1997identifying}, we can control the generated code within a reasonable file size.
Although the size of the generated code is much less than the size of the trace, all communication events, namely MPI function calls and related parameters, are losslessly contained in the code. 
For computation events, we represent each of them by a combination of predefined code blocks.
The combination is determined by solving a constrained quadratic optimization problem, and can accurately approximate the computation performance metrics of a certain execution period of the MPI program.
Such a representation of computation event avoids elaborately profiling the program's computation behavior in instruction granularity that can involve large overhead and interference.

The main contributions of this study are summarized as follows:
\begin{itemize}
\item We propose a novel framework to automatically synthesize proxy-apps for MPI programs. The generated proxy-app can losslessly reproduce the target MPI program's communication behaviors and exhibit similar computation performance metrics.

\item We design an effective approach to extract the structural pattern of the trace and reduce its size by grammar-based analysis. We further transform the compressed trace to executable code as the target proxy-app. 

\item We evaluate our work by representative MPI programs with various settings. The results show that the generated proxy-apps can precisely approximate the performance characteristics of given MPI programs.  

\end{itemize}

The structure of the remaining paper is as follows. 
Section \ref{sec:method} describes the implementation of the proposed proxy-app synthesis framework.
Section \ref{sec:evaluation} shows evaluation results of the proposed method on a series of problem settings.
Section \ref{sec:related} reviews existing studies related to MPI proxy-app synthesis, and makes a brief comparison between this study and theirs.
In section \ref{sec:conclusion}, we draw a conclusion for this study.

\section{Method Implementation}
\label{sec:method}

\begin{figure}[htbp]
\centering
\includegraphics[width = 0.35\textwidth]{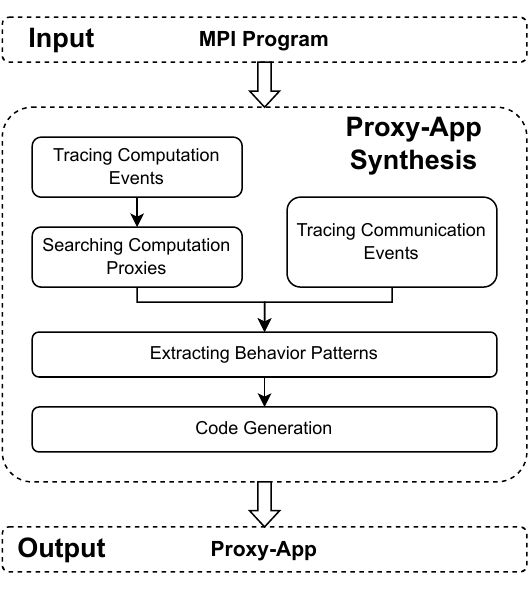}
\caption{Overview of proxy-app synthesis for MPI programs}
\label{workflow}
\end{figure}

\subsection{Overview}
\label{sec:framework}
Figure \ref{workflow} shows an overview of Siesta.  
It takes a binary MPI program as input and finally outputs a C language code as a proxy-app (we can also choose another programming language). The proxy-app's behavior does not accord with the semantics of the input program (e.g., using an algorithm to correctly solve a problem), but has similar performance characteristics.

In the workflow of proxy-app synthesis, 
first, we trace runtime information of communication and computation events.
Second, according to recorded computation characteristics, we search combinations of some predefined code blocks as computation proxies to guide the replay of computation events.
Third, we extract intra-process and inter-process behavior patterns from the trace and obtain a compressed representation of the trace.
Finally, we transform the grammar to C language code as the output.

\subsection{Tracing Communication Events}
Siesta currently supports most synchronous and asynchronous MPI calls, such as MPI\_Irecv, MPI\_Allreduce, etc., with the exception of one-sided communication (e.g., MPI\_Put). 
In addition, I/O trace and MPI-I/O are also not considered in this study.
According to existing studies \cite{tpds16gamblin,ipdps17frank,jpdc19zhang}, the process of I/O trace is similar to that of communication trace. We believe that the method proposed in this study can also be generalized to I/O trace via further engineering efforts.

To trace MPI calls, we use the MPI profiling interface (PMPI) to intercept MPI function calls. PMPI provides convenient interfaces to insert customized code before and after an MPI function. 
Our tracing tool is implemented based on mpiP \cite{ppopp01mpip}, leveraging its higher-level interfaces over PMPI. To capture detailed information for each MPI event, we customized mpiP's implementation of wrapper creations since mpiP only generates aggregated statistics based on all MPI events.

The communication trace we record contains the id of the process (also called rank), name, and parameter information of MPI function calls. 
For data buffers in some functions, like MPI\_Send and MPI\_Recv, we do not record the exact content in the buffer, since it is costly and unnecessary. Recording the data volume is sufficient to reproduce the performance of the MPI function call. 

The values of some handles, like MPI\_Request and MPI\_Comm, are determined at runtime.
Tracing the value change of these handles may prevent many events from being compressed since random values have high information entropy and are difficult to be compressed. 
Besides, to correctly replay the usage of MPI\_Request and MPI\_Comm, recording their exact values is unnecessary.
Our solution is to maintain a pool of free numbers, starting from zero, for tracing these handles.
Taking MPI\_Request as an example, every time a request needs to be used, an unused number is allocated from the pool to the request as its id. When functions such as MPI\_Wait are called, the MPI\_Request object used by MPI library is implicitly released, and then the corresponding number is returned to the request pool. The processing for MPI\_Comm is similar to MPI\_Request.

A process to reduce data redundancy of communication traces is using relative ranks. 
For point-to-point MPI communication functions, the target process is usually related to the process of the function itself.
In other words, for different processes, although their communication target processes are different, the difference between the target process rank and its rank is fixed. 
Especially for mesh-based numerical algorithms in high-performance scientific computing, many processes communicate with their neighbors. 

\subsection{Tracing Computation Events}
Unlike the communication event, the computation event of an MPI program is not a well-defined conception. It is natural to regard an MPI function call as a communication event, while computation behaviors can be defined with many aspects and granularities. 
In this study, we regard all operations between two successive communication events (after the end of the former communication event and before the beginning of the latter communication event) as a computation event. This definition is based on two points. 
(1) Existing studies \cite{sigmetrics15apprime,tc16zhang,jpdc19zhang,tpds21schulz,sc21pilgrim,tpds22sun,tpds23pilgrim} prefer to use the interval between two communication events to indicate a portion of computation time cost. Our definition aligns with the convention.
(2) We can leverage PMPI to trace such computation events, without source code and extra efforts to program slicing and instrumenting. 

To represent communication and computation events with the same trace format, we assume a virtual MPI function that is called MPI\_Compute. 
Each computation event is a call of MPI\_Compute function.
Rather than recording concrete operations in a computation event, which is time-consuming and space-consuming, we take a vector of performance metrics as its footprint.
The parameters of MPI\_Compute record the performance metrics of the corresponding computation event, and replaying a call of MPI\_Compute is to execute a series of operations that exhibit similar performance as the parameters record.
Hence tracing a computation event is to measure its performance metrics.

To obtain performance metrics, we combine our tracing tool with PAPI \cite{browne2000portable} to intercept MPI functions and read hardware performance counters at the same time. 
We select several typical performance metrics as an example, shown in Table \ref{tab:metrics}.
More metrics can also be added with similar operations.
Since the metrics from PAPI are noisy, it is unnecessary to store accurate counts. We set a threshold to cluster similar computation events into one event so that the number of unique calls of MPI\_Compute can be reduced.

\begin{figure*}[!htbp]
\centering
\includegraphics[width = 0.85\textwidth]{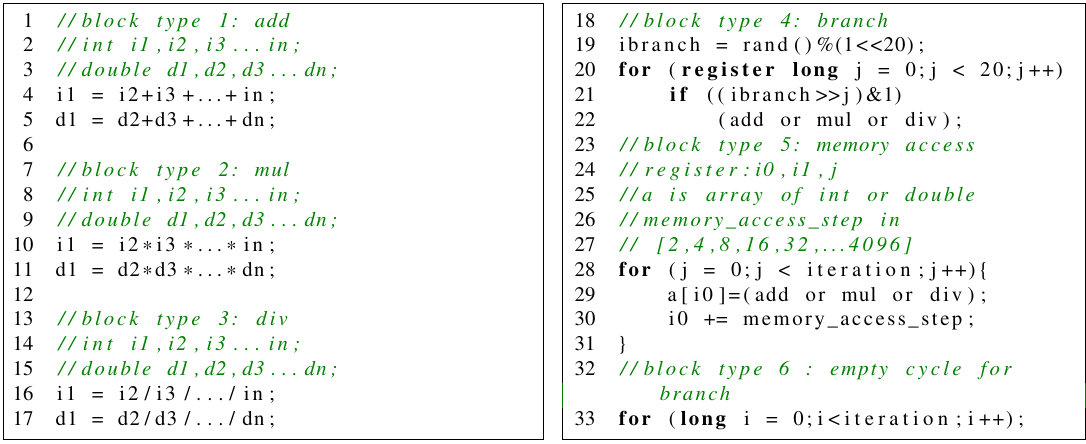}
\caption{Code block examples for mimicing behaviour of application}
\label{basicblocks3}
\end{figure*}

\subsection{Searching Computation Proxies}
\label{sec:search}
To generate a compilable and executable proxy for a computation event, we need to infer its possible source code according to obtained performance metrics. 
There might be numerous possible codes that show similar performance metrics.
To make a simplified approximation, we define the search space by linear combinations of some pre-designed code blocks that show diverse performance. 
Searching for an optimal combination with maximized performance similarity (namely minimized error) can be formulated as a constrained optimization problem.

Generally, for a given computation event, we can obtain its $n$ performance metrics $\mathbf{t}\in \mathbb{R}^{n\times 1}$. These metrics constitute the footprint of a computation event and can be used to measure the similarity of different computation events.
Suppose there are $m$ code blocks called $block_{1,2,...,m}$ and we can use micro-benchmarks to get the $i$-th metric of $block_j$, called $b_{i,j}$ on a given HPC system. Let 
    $\mathbf{B} = (b_{i,j})_{n\times m}$.
\begin{table}[t]
    \centering
    \caption{Performance metrics}
    \begin{tabular}{cll}
        \toprule
         Id&Name&Description\\
         \midrule
         1&INS& Instructions \\
         2&CYC& Cycles\\
         3&LST& Load/store \\
         4&L1\_DCM& L1 data cache misses\\
         5&BR\_CN& Conditional branch\\
         6&MSP& Mispredicted conditional branch\\
         7&DTLB\_LOADS& Data TLB Loads \\
         8&FP\_ARITH& Float Point Arithmetic Count \\
         \bottomrule
    \end{tabular}
    \label{tab:metrics}
\end{table}
Then we combine the $m$ code blocks linearly to obtain a computation proxy. Suppose $block_j$ is repeated $x_j$ times and let $\mathbf{x} = (x_1,x_2,...,x_m)^\top\in \mathbb{R}^{m\times 1}$. We will make a rounded approximation at the end because $\mathbf{x}$ must be positive integers. At this point, the original problem can be transformed into a constrained optimization problem as follows,
\begin{align}
    \begin{aligned}
    \min_{\mathbf{x}} &f(\mathbf{x}) = \|\mathbf{Bx-t}\|^2,\\
    &x_i\geq 0 ,\quad i=1,2,...,m
    \end{aligned}
\end{align}
Since the order of magnitude difference between different metrics may be large, we divide matrix $B$ by rows and changed the difference of each metric to a relative error. 
\begin{equation}
    \mathbf{B} = (b_{i,j})_{n\times m}=(\mathbf{b}_{1},\mathbf{b}_{2},...,\mathbf{b}_n)^\top ,\mathbf{b}_i\in \mathbb{R}^{1\times m}.
\end{equation}
The modified optimization problem is:
\begin{align}
\begin{aligned}
    \min_{\mathbf{x}} &f(\mathbf{x}) = \sum_{i=1}^n\frac{1}{t_i^2}(\mathbf{b}_i\cdot \mathbf{x}-t_i)^2,\\
    &x_i\geq 0 ,\quad i=1,2,...,m\label{proxy:equation}
\end{aligned}
\end{align}

We select eight performance metrics as an example, detailed in Table \ref{tab:metrics}, to represent various aspects of a computation event's behaviors. 
By comparing these metrics we can assess the similarity between the computation event and our synthesized computation proxy.
These metrics are commonly employed in performance evaluation, and we can expect that the code synthesized accordingly will have similar characteristics to the computation event.

We have developed a series of code blocks, illustrated in Figure \ref{basicblocks3}, to approximate various aspects of application behaviors, including computation, branching, and memory access. 
These blocks exhibit distinct performance characteristics, enabling effective manipulation of specific behaviors by adjusting the repetition count of each block. 
Using Equation \ref{proxy:equation}, we can determine the optimal linear combination of these code blocks to obtain the computation proxy. It is a convex quadratic optimization problem, and existing algorithms and libraries are already well-suited to solve it.

\subsection{Extracting Intra-process Behavior Patterns}
HPC applications conduct iterative communications and computations to solve problems.
Consequently, their traces contain repetitive data.
We adopt a grammar-based method \cite{nevill1997identifying} to extract hierarchical and periodic patterns of nested loops in a trace and compactly represent the patterns with small space consumption.

For representing symbolic sequences, context-free grammars have been widely used in text compression, natural language processing, music processing, and macromolecular sequence modeling \cite{dorier2015using}. 
Context-free grammar provides a simple and mathematically precise mechanism for describing the methods by which phrases in language are built from smaller blocks, capturing the block structure of sentences in a natural way. In a trace, this block structure can be used to describe loop sections or recursive content.

Before the extraction, events are stored in a hash table. A key in the hash table is a string that records the information of a unique event, and the corresponding value is an event id.
Then the trace is represented by a sequence of ids.
Each process of the MPI program has its own trace and table, so the extraction consists of two phases: extracting intra-process patterns of each process respectively, and further extracting inter-process across all processes.


Specifically, we adopt Nevill's Sequitur algorithm \cite{nevill1997identifying} to convert a trace to a context-free grammar with linear worst-case time and space complexity. It only needs to scan the trace once to complete the construction of the context-free grammar. The algorithm starts with an initial symbol $S$, which is also called \textbf{main rule}. Whenever a new terminal $x$ is encountered, it adds $x$ to the last position of $S$, and then recursively enforces two constraints:
\begin{enumerate}
\item 
A symbol pair consisting of any two adjacent symbols $ab$ occurs only once in the entire grammar. If there is a duplication, generate a new rule, $A \rightarrow ab $, and enforce this constraint recursively.
\item 
All rules (except the main rule) will be applied at least twice. If a rule is applied only once, replace the non-terminal with the content of the rule where the non-terminal corresponding to the rule appears, remove the rule, and enforce the constraint recursively.
\end{enumerate}

The size of the context-free grammar generated by the sequitur algorithm is determined by the input sequence. In the best case, the size of such a grammar grows logarithmically when the symbols in the input sequence appear periodically. For example, a sequence $aaaaaaaa$ can be represented by the following grammar:
\begin{equation} 
S \rightarrow AA, A \rightarrow BB, B \rightarrow aa.
\end{equation}
For sequences containing regular loop structures, the optimization scheme \cite{dorier2015using} can reduce the space occupied by the generated grammar from $O(log(n))$ to $O(1)$. The optimization method  adds an attribute to all symbols in the grammar, which marks the repetition times of the corresponding symbol, that is, for a symbol $a$ repeated $i$ times, it is represented by $a^i$. Meanwhile, we add a constraint to the grammar generation process:
\begin{enumerate}
\setcounter{enumi}{2}
\item 
Any two adjacent symbols $a^i$, $a^j$ will be converted to $a^{i+j}$, recursively enforcing this constraint.
\end{enumerate}

\subsection{Extracting Inter-process Behavior Patterns}

If all processes keep a local symbol table, as the number of processes increases, the size of the resulting grammar will grow linearly. Therefore, we need to merge the symbol table across processes. Due to the wide use of the SPMD programming style in MPI programs, similar MPI functions are often called between processes. In collective communication, all processes share the exact same MPI functions. In addition, many inter-process point-to-point communications often exhibit similar behavior with the encoding method.

\subsubsection{Merging Terminal Tables}

Although the number of MPI function calls increases linearly with the number of processes, many MPI programs exhibit a significant amount of duplication in terminals between processes, which can be eliminated by recording the repeated terminals once and assigning a unique global number.
The global id for computation terminals has already been generated and can be used during the merging process between processes.
The time complexity of the entire merging process is $log_2P$, where $P$ is the number of processes. After the entire merge process is complete, the root process broadcasts a table of terminals to all other processes.

\begin{figure}[!htbp]
\centering
\includegraphics[width = 0.48\textwidth]{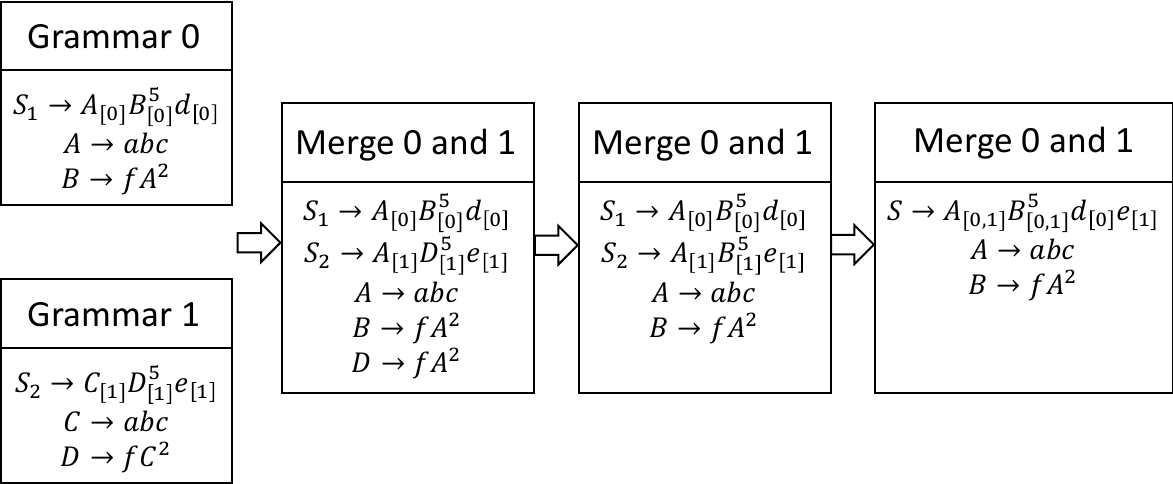}
\caption{Example of merging two grammars}
\label{grammar_merge}
\end{figure}

\subsubsection{Merging Non-terminal Tables}
To avoid conflicts, each terminal is substituted with a globally unique id, and each non-terminal is also uniquely identified. Identical non-terminals from different processes are merged based on depth, which measures the tree height in a tree representation where non-terminals are non-leaf nodes and terminals are leaf nodes. Merging is only possible for non-terminals with the same depth. We prioritize merging shallower trees first, sorting non-terminals by depth to minimize symbol table scans.

The main rule, representing a process's overall logic, is often longer than other rules and exhibits similarities across processes due to the Single Program Multiple Data (SPMD) nature of many MPI programs. For merging main rules, we employ a longest common subsequence (LCS) approach. Each symbol in the main rule is initially tagged with its process number. We then compute the LCS between the main rules of two processes, merge the rank lists for symbols in the main rule, and maintain the original order for other symbols, as demonstrated in Figure \ref{grammar_merge}.

Merging of main rules is limited to processes with substantial similarity. Merging dissimilar main rules can result in a rule longer than the combined original rules, leading to excessive branching in the subsequent code generation. Thus, the main rules are first clustered based on minimum edit distance, and then rules within the same group are merged.

\subsection{Code Generation}

        

\begin{figure}[!htbp]
\centering
\includegraphics[width = 0.48\textwidth]{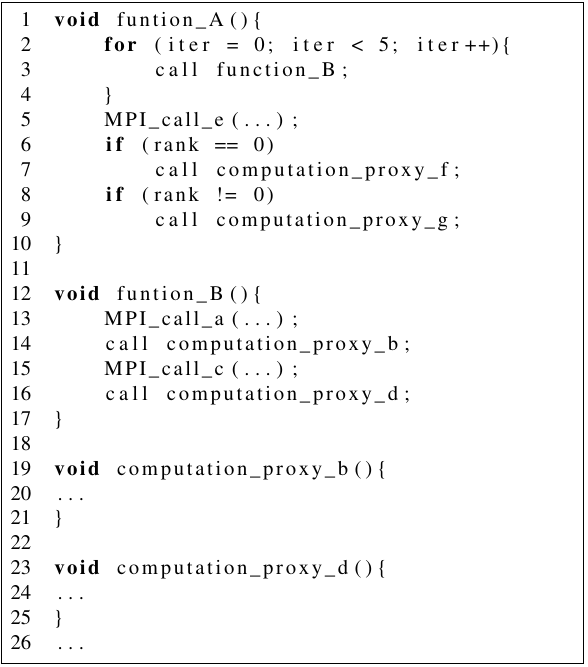}
\caption{Code generation example with grammar in Equation \ref{code_gen:grammar}}
\label{code_gen:code}
\end{figure}

By extracting intra-process and inter-process patterns, we can generate a symbol table and a grammar. By looking up the symbol table and traversing the grammar, we then produce executable C code. We use the grammar \ref{code_gen:grammar} as an example and generate the code snippet in Figure \ref{code_gen:code}. This grammar contains non-terminals $S,A,B$, communication events $a,c,e$ and computation events $b,d,f,g$.
\begin{align}
\label{code_gen:grammar}
    S&\rightarrow A \notag \\
    A&\rightarrow B^5ef_{[0]}g_{[1-n]}\\
    B&\rightarrow abcd \notag
\end{align}

Initially, we convert a symbol table containing terminals and non-terminals into C code. Terminals, which include communication and computation events, are translated to MPI functions and computation function calls respectively, using parameter information  and computation proxies. This corresponds to Line 13-16, 19-25, etc. of Figure \ref{code_gen:code}.

Non-terminals are processed recursively by treating each as a function composed of terminals and other non-terminals. This involves generating function calls for terminals and direct calls for non-terminal functions within the content of a non-terminal function, like $function\_A$ and $function\_B$ in Figure \ref{code_gen:code}.

The main rule of a process includes process ranks for each symbol within its function content, dictating which processes execute the corresponding function. Conversion of the main rule into C code involves creating branch statements based on these ranks to guide process execution and merging consecutive symbols with identical ranks to minimize redundant branches.

\begin{table}[!htbp]
    \centering
    \caption{Platform specification}
    \begin{tabular}{lll}
        \toprule
         &Platform A&Platform B\\
         \midrule
         Processor   &\makecell[l]{Intel Xeon \\Scale 6248}&\makecell[l]{Intel Xeon \\Gold 6258R}\\
         \# Cores/node      &$20\times 2$   &$28\times 2$ \\
         Memory             & 192 GB        & 192 GB \\
         L1 I/D             &32 KB          &32 KB\\
         L2                 &1024 KB        &1024 KB\\
         Frequency          &2.5 GHz        &2.7 GHz\\
         \bottomrule
    \end{tabular}
    \label{tab:platforms}
\end{table}

\begin{table}[!htbp]   
    \caption{Specification of Generated proxy-apps}
    \centering
    \begin{tabular}{crrrr}
        \toprule
         Program&Process&Trace size&Overhead&Error\\
         \midrule
         \multirow{4}{*}{BT}    
             & 64  & 290 MB     & \textless 1\% & 0.39\% \\
             & 121 & 767 MB     & \textless 1\% & 0.50\% \\
             & 256 & 2,367 MB   & \textless 1\% & 1.07\% \\
             & 529 & 6,962 MB   & \textless 1\% & 1.04\% \\
         \midrule
         \multirow{4}{*}{CG}    
             & 64  & 491 MB     & 2.2\%    & 4.45\% \\
             & 128 & 1,255 MB   & 2.26\%    & 2.40\% \\
             & 256 & 2,534 MB   & 5.64\%    & 2.09\% \\
             & 512 & 6,223 MB   & 5.34\%    &1.57\% \\
         \midrule
         \multirow{4}{*}{SP}
             & 64  & 508 MB    & \textless 1\%      & 0.46\% \\
             & 121 & 1,317 MB  & 2.04\%    & 0.84\% \\
             & 256 & 4,024 MB  & 3.21\%    & 0.50\% \\
             & 529 & 11,662 MB & \textless 1\%      & 1.12\% \\
         \midrule
         \multirow{4}{*}{EP}    
             & 64  & 32 KB     & \textless 1\%  & 3.64\% \\
             & 128 & 64 KB     & \textless 1\%  & 2.74\% \\
             & 256 & 128 KB    & 4.15\%    & 3.24\% \\
             & 512 & 256 KB    & 3.93\%    & 2.23\% \\
         \midrule
         \multirow{4}{*}{MG}    
             & 64  & 168 MB     & \textless 1\%      & 0.36\% \\
             & 128 & 318 MB     & \textless 1\%      & 0.65\% \\
             & 256 & 618 MB     & 3.07\%    & 0.72\% \\
             & 512 & 1,294 MB   & 7.82\%    & 0.85\% \\
         \midrule
         \multirow{4}{*}{Sweep3d}
             & 64  & 619 MB     & 1.65\%            & 1.17\% \\
             & 128 & 1.3 GB     & \textless 1\%     & 1.34\% \\
             & 256 & 3.3 GB     & \textless 1\%     & 2.73\% \\
             &512  &6.6 GB      & \textless 1\%     &2.99\%\\
         \midrule
         \multirow{4}{*}{Sod}   
             & 64  & 6 MB      & \textless 1\% & 1.97\% \\
             & 128 & 12 MB     & \textless 1\% & 3.71\% \\
             & 256 & 24 MB     & \textless 1\% & 4.27\% \\
             & 512 & 48 MB     & \textless 1\% & 3.07\% \\
        \midrule
         \multirow{4}{*}{StirTurb} 
             & 64  & 304 MB    & \textless 1\% & 3.92\% \\
             & 128 & 614 MB    & \textless 1\% & 3.71\% \\
             & 256 & 1,241 MB  & \textless 1\% & 8.67\% \\
             & 512 & 2,510 MB  & \textless 1\% & 5.37\% \\
         \midrule
         \multirow{4}{*}{Sedov} 
             & 64  & 16 MB      & \textless 1\% & 2.90\% \\
             & 128 & 32 MB      & \textless 1\% & 5.38\% \\
             & 256 & 64 MB      & \textless 1\% & 3.56\% \\
             & 512 & 135 MB     & \textless 1\% & 3.49\% \\       
         \bottomrule
    \end{tabular}
    \label{tab:Specification_of_Generated_proxy-apps}
\end{table}

\section{Evaluation}
\label{sec:evaluation}
\subsection{Setup}
We select a variety of MPI programs as targets for evaluation, as shown in Table \ref{tab:Specification_of_Generated_proxy-apps}. BT (block tridiagonal), CG (conjugate gradient), SP (scalar pentadiagonal), EP (embarrassingly parallel), and MG (multi-grid) are kernels or pseudo applications from the NAS Parallel Benchmark (NPB) \cite{bailey1991parallel}. SWEEP3D \cite{yoon2007productivity} addresses the neutron transport problem in discrete coordinate 3D Cartesian geometry.
FLASH \cite{fryxell2000flash} is a production software package for handling general compressible flow problems found in many plasma physics experiments and astrophysical environments. We select three representative scientific simulation problems, Sod, Stirturb, and Sedov, to evaluate FLASH.

The problem size of the NPB programs we evaluate is D scale on Platform A and E scale on Platform B, and the version is NPB 3.3.1. The input size for all FLASH problems is 64*64*64. For SWEEP3D, we use an input of scale 1000*1000*1000. All evaluated programs are compiled with GCC 4.8.5 and OpenMPI 3.1.0. We implement a tracing tool based on mpiP 3.5 to generate traces for these programs.
The experiments are conducted on two platforms, as shown in Table \ref{tab:platforms}. 

\begin{figure*}[htbp]
\centerline{\includegraphics[width = 0.80\textwidth]{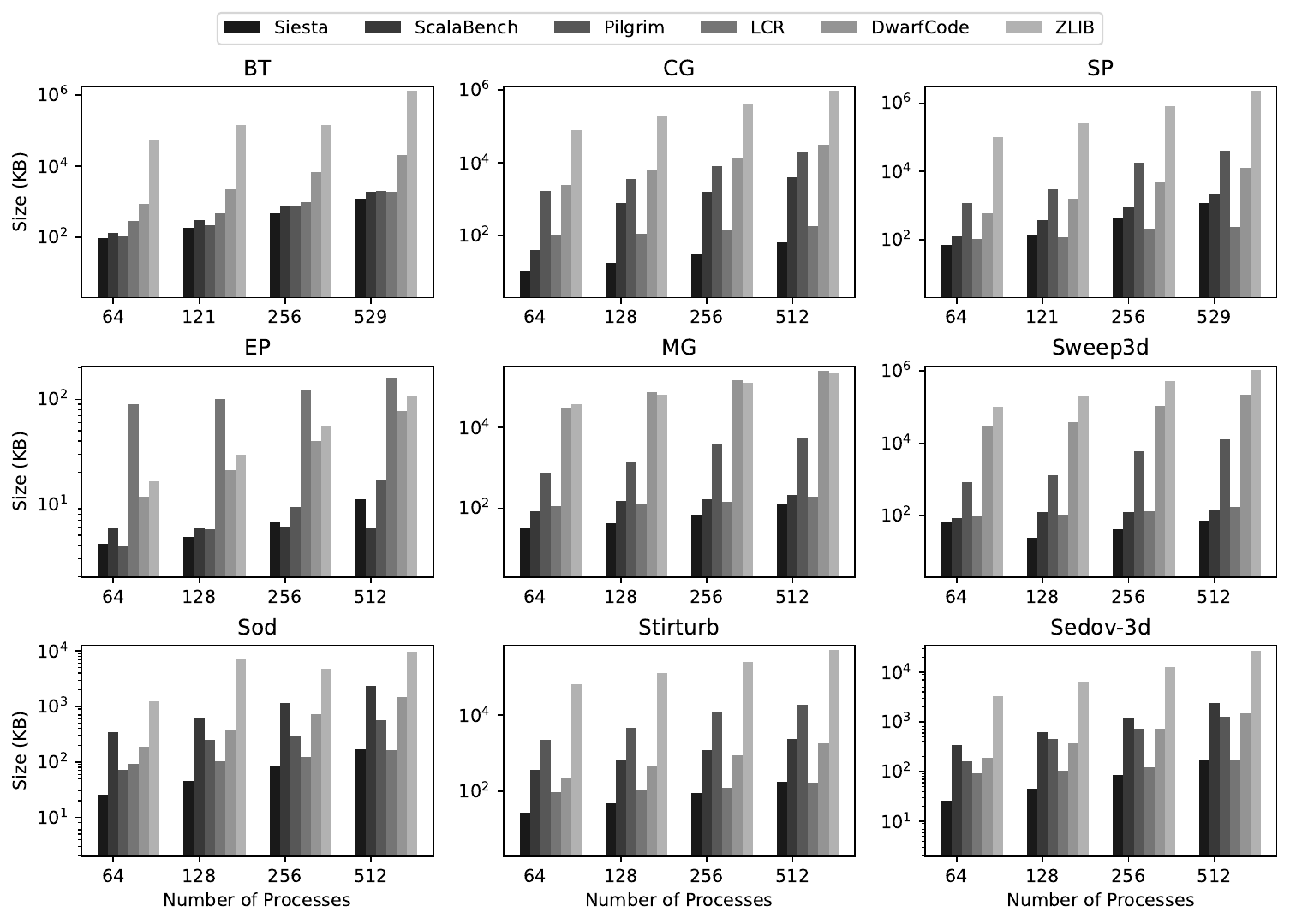}}
\caption{Comparison of compressed data size between our method and other methods}
\label{size_compare}
\end{figure*}

\subsection{Specification of Generated Proxy-apps}
Table \ref{tab:Specification_of_Generated_proxy-apps} shows the specification of generated proxy-apps in our evaluation.
The trace size column shows the total size of the original trace files generated by each program without compression. 
The overhead column indicates the impact of tracing instrumentation on the execution of the original program. 
The overhead is calculated by the relative difference between the time costs of executing MPI programs with and without instrumentation. 
It can be found that, for evaluated programs, although we record information about hardware performance counters in addition to intercepting the MPI functions, both of them bring a small overhead to program execution. 
It is to ensure that the tracing does not interfere with the original pattern of the target program a lot.

To measure the similarity between the original MPI program and the generated proxy-app, we calculate the relative error between their performance metric values. 
The relative error is calculated by the absolute difference between the metric values divided by the original program's metric value. 
We then compute the average relative error across all the metrics and processes to measure the overall error between the two programs.
As shown in Table \ref{tab:Specification_of_Generated_proxy-apps}, the discrepancy between the proxy generated by Siesta and the original MPI program is minor, with an average error rate of 2.48\% across all tested programs and processes. Note that the proxies generated by other methods, such as ScalaBench and Pilgrim, do not reproduce the original application's computation behaviors. Therefore, determining an error for them in this context is not feasible.

\subsection{Comparison of compression ability}
Compression is the key process to extract behavior patterns of MPI programs from a large volume of trace data.
This section presents a comparison of the compressed data size achieved by Siesta in contrast to other trace compression methods, including ScalaBench \cite{ipdps12frank}, Pilgrim \cite{sc21pilgrim}, \cite{tpds23pilgrim}, LCR \cite{tpds22sun}, DwarfCode \cite{tc16zhang}. 
These methods focus on compressing communication traces while only recording the execution time for computational events. 
We also take a domain-agnostic compressor ZLIB as a baseline, which is used in the Open Trace Format \cite{iccs06wolfgang}.
Directly comparing Siesta's compressed data size with theirs does not yield a fair assessment, since Siesta records more detailed computation metrics. 
To ensure fairness in comparing compression ability, we combine Siesta with a histogram-based method, which is also used in ScalaBench, to aggressively compress computational metrics. 

Figure \ref{size_compare} illustrates the comparison of compressed trace sizes among the evaluated methods. 
ZLIB has the largest file sizes, due to its unawareness of the holistic structure of the trace data.
In the task-specific compression methods, Siesta and LCR achieve the smallest file sizes in general.
Siesta has better performance than LCR method for CG, MG, EP, and Sweep3d. 
However, LCR exhibits better scalability as the number of processes increases. 
This advantage stems from that LCR models communication events as a random process, while our approach remains lossless in recording the communication event sequence.


\begin{figure*}[!htbp]
\centering
\includegraphics[width = 0.8\textwidth]{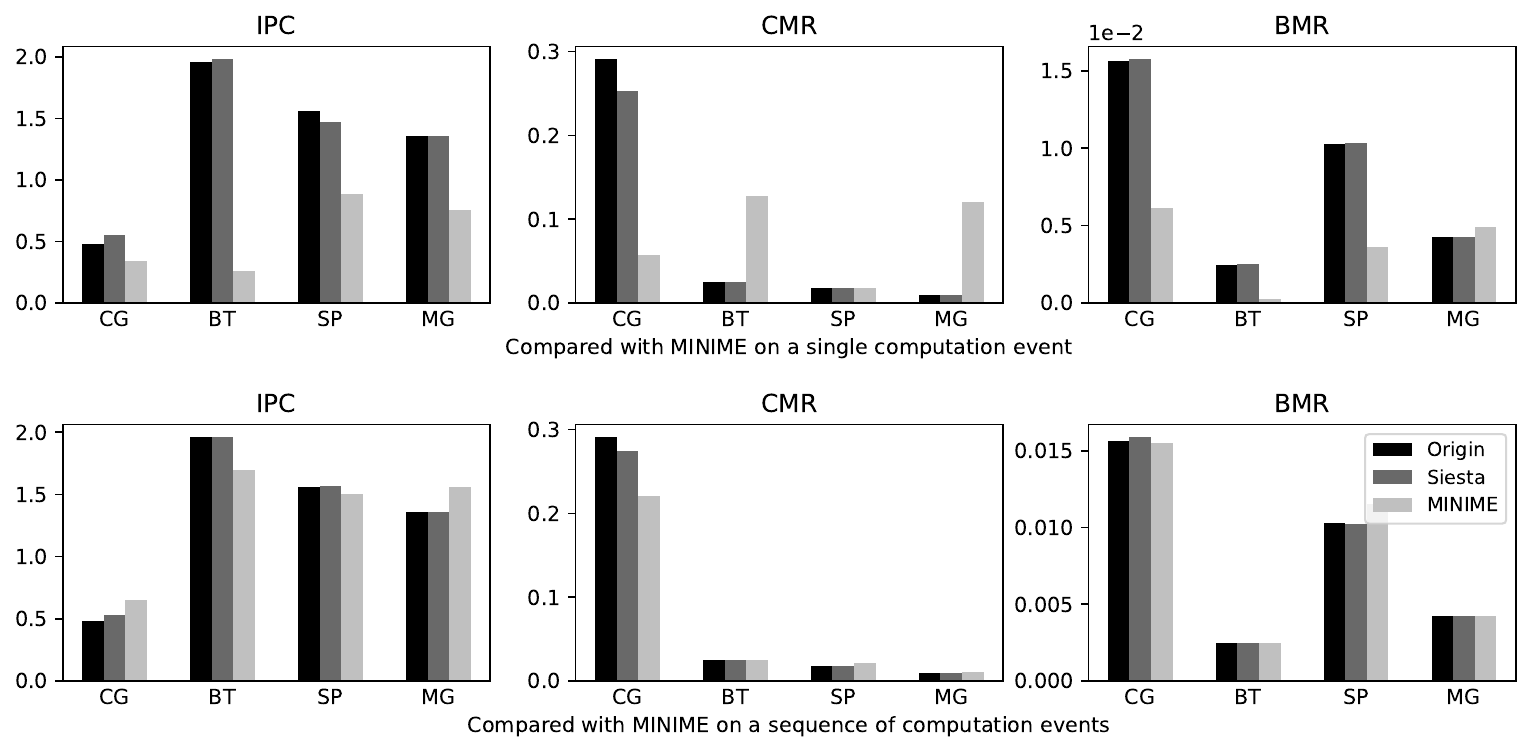}
\caption{Comparison of proxy-app performance metrics}
\label{minime}
\end{figure*}

\subsection{Comparison to MINIME}
To demonstrate the effect of mimicking performance characteristics, we compare Siesta with MINIME \cite{tc15minime, date17minime}. MINIME uses an iterative approach to adjust code blocks and mimic the performance of the original program. 
It uses Instructions Per Cycle (IPC), Cache Miss Rate (CMR), and Branch Misprediction Rate (BMR) to measure performance similarity to the original program, so we also use these three metrics for comparison.

We first experiment with a single computation event, as shown in the top three figures in Figure \ref{minime}. The origin in the figure corresponds to the sum of the computational parts of the tested programs.  
We consider the whole program execution as a single computation event and mimic it by MINIME and Siesta. 
For example, we sum up all the instructions and cycles of the computation part of a tested program and get its IPC.
It can be seen that, for a single computation event, Siesta works better than MINIME. 
The difference stems from that Siesta introduces new code blocks with diverse performance characteristics, so that their combination can approximate an extensive range of performance metrics. 
Besides, MINIME models the problem as a multi-objective optimization task and solves it using a local search strategy, which may lead to a local optimum.

To elaborately mimic the performance characteristics of an MPI program in different execution periods, it is necessary to regard the execution as a sequence of events and mimic each of them separately. 
As shown in the bottom three figures in Figure \ref{minime}, when we mimic each computation event separately and sum them up, Siesta shows much higher similarity to the original program compared to MINIME.

\subsection{Comparison to ScalaBench}
\subsubsection{Time span similarity}
In this section, we compare the results of Siesta and ScalaBench on Platform A, in terms of the error of reproducing the original program's execution time.
Pilgrim is not taken as a baseline, although it is an advanced MPI trace compression tool and provides proxy-app generation functionality.
Due to it only focuses on compressing and replaying communication information, without filling in the execution time of the computation part, the execution results of Pilgrim cannot reflect the original program's execution time, and the average error of its execution time is 84.30\% across all programs we tested. 
The large error indicates that Pilgrim may not generate proxy-apps with thorough performance approximation, so we mainly compare Siesta with ScalaBench in later experiments.

\begin{figure*}[ht]
\centering
\includegraphics[width = 0.78\textwidth]{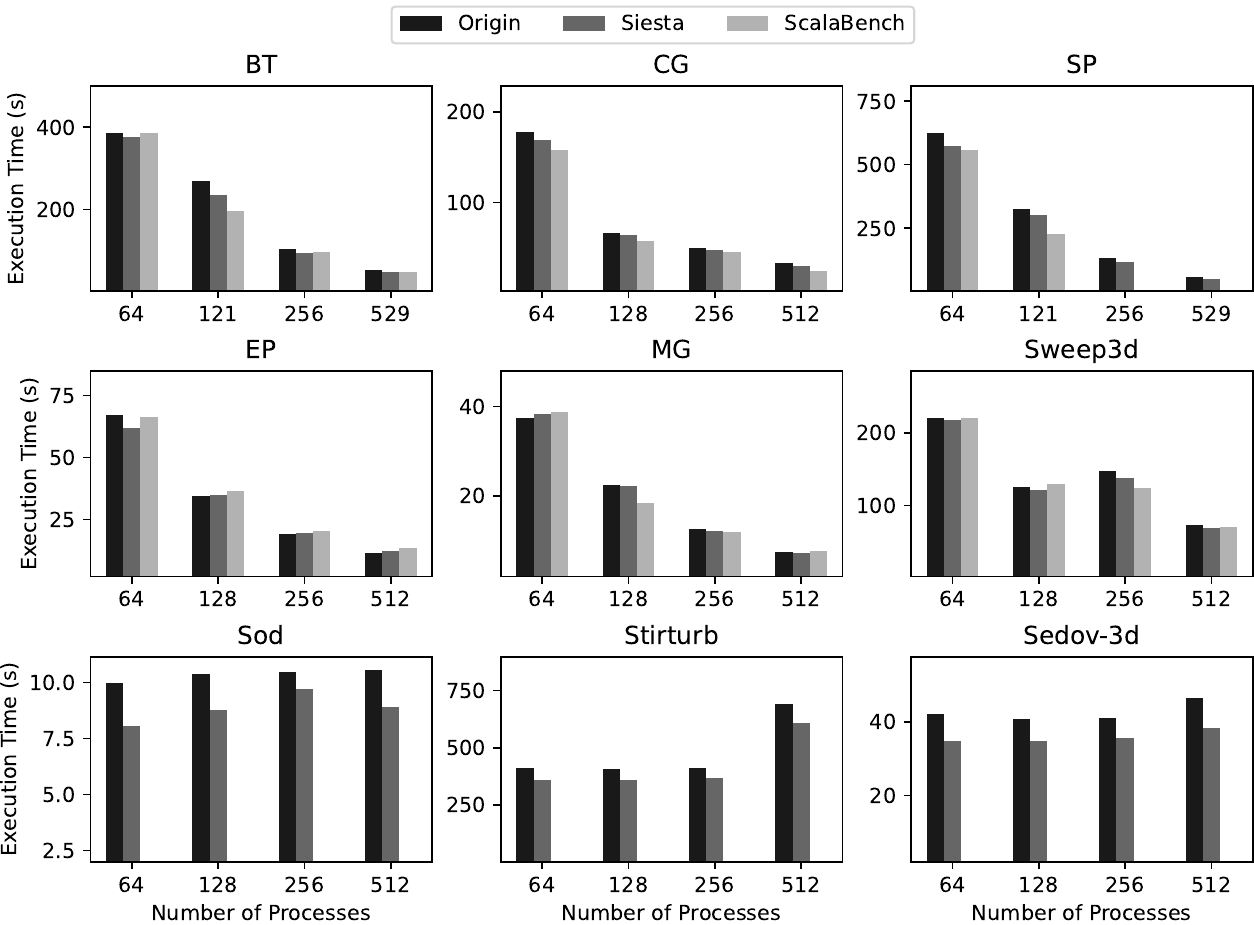}
\caption{Comparison of proxy-app execution time}
\label{execution_time}
\end{figure*}

\begin{figure*}[!tp]
\centering
\includegraphics[width = 0.78\textwidth]{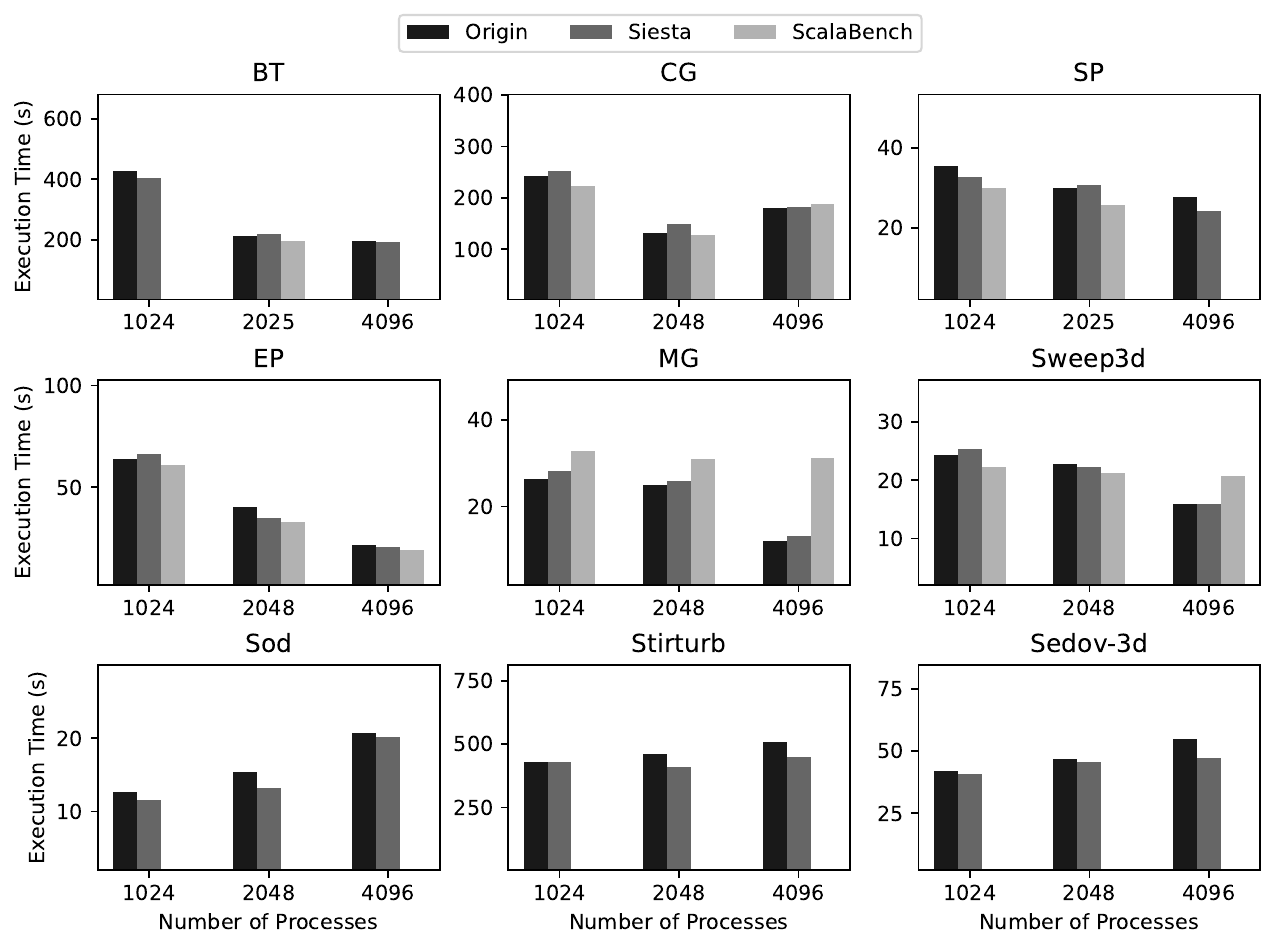}
\caption{Comparison of proxy-app execution time at large scale}
\label{large-scale}
\end{figure*}

Figure \ref{execution_time} shows the execution times of evaluated proxy-apps and the original programs. 
Siesta has similar execution time to the origin and outperforms Scalabench in the majority of cases.
The mean percentage error across all programs  can be calculated by the formula
$100\% \times  \left| (T_{gen}-T_{app})/T_{app} \right|$.
The errors of Siesta and ScalaBench are 5.01\% and 10.22\%, respectively.
ScalaBench crashed when generating executable proxy-apps for certain programs, such as SP with 256 and 529 processes, and the three scientific simulations in the FLASH package. Therefore, we are not able to show the execution time of ScalaBench for these programs.
Siesta represents the trace compactly while still preserving important details. 
Our approach can record more information compared to ScalaBench and is applicable to a wider range of programs.

Due to the limitation of the number of machines on Platform A, we tested large-scale experimental results of Siesta on Platform B. The
experimental results are shown in Figure \ref{large-scale}.
As the number of processes used increases, the proportion of communication time will increase. 
ScalaBench crashed when dealing with FLASH applications and BT with 1024 and 4096 processes.

In large-scale experiments, most tested applications are not computation-bound but communication-bound, so the execution time of the MPI program may not scale with process numbers. Since ScalaBench does not hold the same communication structure as the original MPI program, the execution time of its proxy becomes unfaithful, especially in MG. In contrast, the proxy generated by Siesta still maintains a high similarity in execution time compared to the original program.
The mean percentage errors of Siesta and ScalaBench are 5.62\% and 25.53\%, respectively.

\begin{figure*}[!tp]
\centering
\includegraphics[width = 0.8\textwidth]{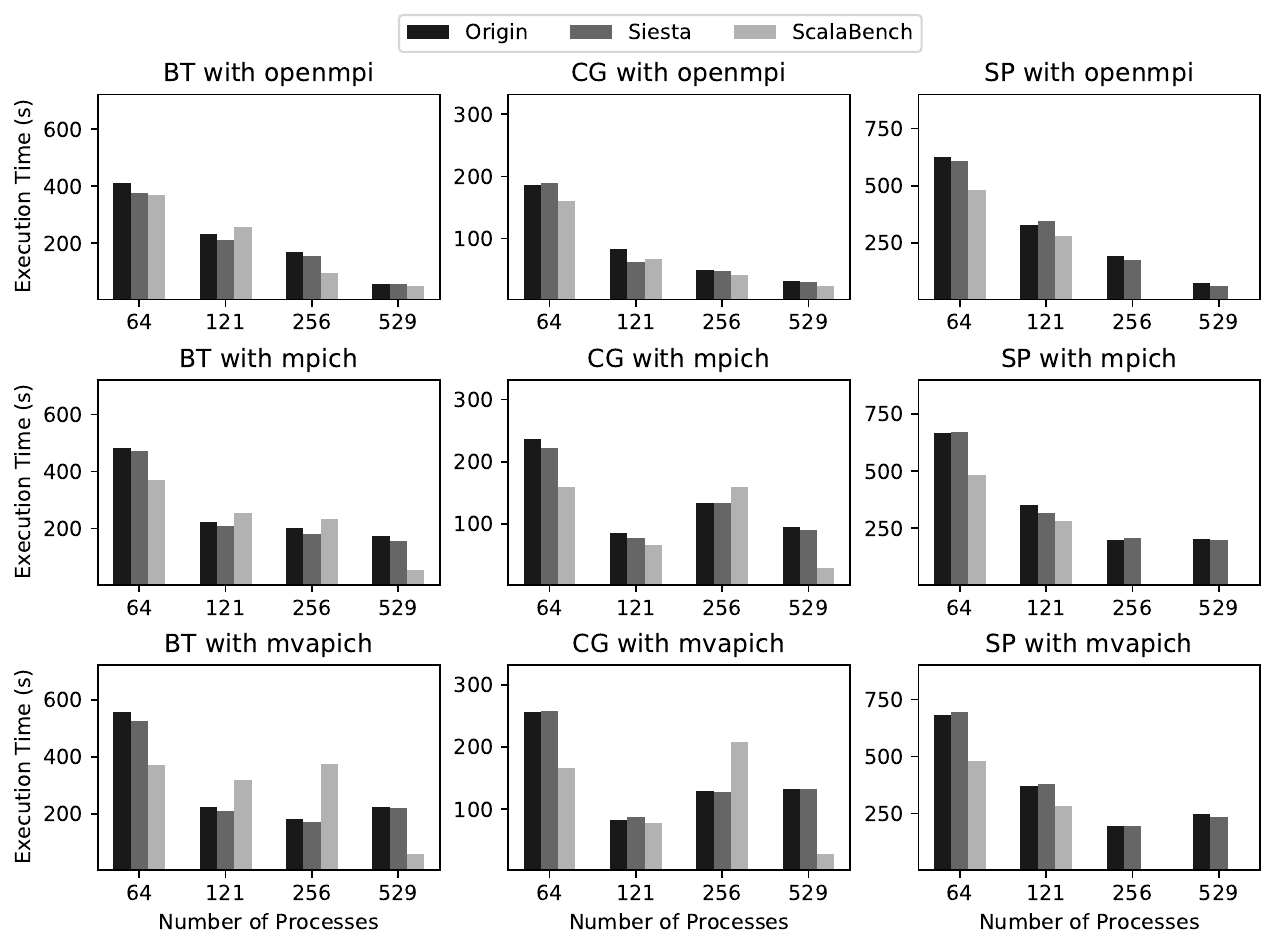}
\caption{Comparison of proxy-app execution time using different MPI implementations}
\label{compiler}
\end{figure*}




\subsubsection{Robustness to MPI Implementation Changes}

In this section, we compare the execution time accuracy with ScalaBench under different MPI implementations. The experimental results are shown in Figure \ref{compiler}. As mentioned above, ScalaBench lacks data for SP under 256 and 529 processes and FLASH programs. 
The proxy-apps in the experiments are generated under openmpi.
We then execute the proxy-apps using openmpi, mpich and mvapich, respectively. 

In different MPI implementations, due to the different support for shared memory and network, the performance of MPI communication functions is also different, which has a direct impact on program execution time. Because Siesta preserves the communication mode in the original program trace losslessly, it can reproduce the original program execution time well with different MPI implementations.
Specifically, for Siesta, the mean percentage error is 5.78\%, while
the error for ScalaBench is 33.58\%.
In ScalaBench, the iterative matching criteria is relaxed during compression, and a statistical histogram is constructed to record some parameter information, so that the communication mode of the original program cannot be completely restored when the proxy-app is generated. 
For example, when sending functions and receiving functions cannot match, ScalaBench generates additional MPI\_Irecv functions to match the MPI\_Send function, which are not the original behaviors of target MPI programs. Therefore, when changing the MPI implementation, ScalaBench may not reproduce the execution time of the original program precisely.

\section{Related Work}
\label{sec:related}
\subsection{Trace Compression}
Tracing MPI programs' execution on HPC systems is a basic approach for performance analysis. Typical applications of trace include visualization \cite{jss18loic}, debugging \cite{pldi15light,arxiv20debug},  performance estimation \cite{tpds17astro}, inefficiency mining \cite{tpds16gamblin,sc20zhai}, benchmark or proxy-app synthesis \cite{ipdps12frank,sigmetrics15apprime,tc16zhang,jpdc19zhang}, etc.
Due to the increasing complexity and scale of HPC applications and systems, 
the storage cost of traces is becoming unaffordable.

Open trace format (OTF) \cite{iccs06wolfgang} and its variant \cite{eurompi15otf} are standard trace formats supported by many trace tools, like Vampir \cite{pthpc08vampir} and Tau \cite{parco11tau}. OTF adopts zlib \cite{rfc96zlib} to compress traces. Since zlib is agnostic to the repetition structure among traces, the compression ratio can be further improved.

Effective trace compression methods adopt specific data structures and matching algorithms to extract the repetition patterns in traces.
ScalaTrace \cite{jpdc09mueller} and its successors \cite{icpp11frank, ppopp11frank, ics13scalatraceii, ipdps17frank,jpdc17frank, ipdps18chameleon} use data structures called regular section descriptors (RSD) and power-RSD to greedily find repetitive sequences in a trace.
DwarfCode \cite{tc16zhang,jpdc19zhang} models the trace compression problem as searching an optimal combination of primitive and inextensible tandem arrays, which is an NP-hard problem. To solve the problem in reasonable time complexity, DwarfCode adopts suffix trees to find a suboptimal solution to represent the repetition patterns of given traces.
Omnisc'IO \cite{dorier2015using} and Pilgrim \cite{sc21pilgrim, tpds23pilgrim} use context-free grammars to represent traces. They use a modified Sequitur algorithm to generate the grammars, so that the size of grammars can be effectively controlled.
Zhai et al. \cite{sc14zhai,atc18zhai,ppopp20zhai,sc20zhai} conduct static top-down analysis on MPI program's source code, so they can efficiently obtain the repetition patterns. Note that the source code is not only from the program, but also from its dependent libraries. The requirement of the source code limits the application scenario of their method, especially when considering proprietary or confidential programs. 

\subsection{Proxy-app Synthesis}
A trace compression method can be used to generate synthetic proxy-apps (also called benchmarks in some literature), if the compressed trace can be transformed into executable code.
ScalaBench \cite{ipdps12frank} is a proxy-app synthesizer based on ScalaTrace variants. 
It uses histograms to approximate the communication parameter distributions of MPI programs. To ensure the approximated communication can be smoothly replayed, without mismatch or deadlock, it also implements a coordinator to control communication behaviors at runtime.
DwarfCode outputs executable code as the result of trace compression, so it is naturally a proxy-app synthesizer.
APPRIME \cite{sigmetrics15apprime} focuses on synthesizing proxy-apps to represent iterative simulation workloads for scientific computing.
Astro \cite{tpds17astro} focuses on synthesizing proxy-apps to estimate the performance of MPI programs on larger parallel scales.
Both APPRIME and Astro adopt a stochastic Markov process to model the status transition of MPI programs.

Kim et al. \cite{iiswc14kim}, Panda et al. \cite{pact17lizy}, and MINIME \cite{tc15minime,date17minime} investigate synthesizing proxy-apps that fit given computation characteristics in certain scenarios, by designing and combining a series of code blocks. Kim et al. focus on APPs running on ARM-based mobile devices. Panda et al. focus on database workload.
MINIME focuses on multi-thread programs with predefined parallel patterns.
Since they only concern about shared-memory programs on a single node, their works do not involve communication events and do not need to analyze traces.

\subsection{Comparison}
The design of Siesta is inspired by many existing studies.
Here is a summary of the difference between Siesta and related studies.
Compared with ScalaBench, APPRIME, DwarfCode, and Pilgrim, Siesta achieves lossless representation of communication events and high-fidelity approximation on computation characteristics, while they conduct lossy compression on communication events and only regard computation behaviors as time intervals.  
Besides, we improve the extraction of inter-process patterns, achieving better compression than the state-of-the-art compression method Pilgrim.
Compared with Kim's work \cite{iiswc14kim}, Panda's work \cite{pact17lizy}, and MINIME \cite{tc15minime,date17minime}, Siesta solves a more general and challenging computation proxy synthesis problem on distributed architectures, where computation behaviors are more diverse than that on shared-memory architectures they concern. 
We also propose a new design of predefined code blocks and a combination search algorithm that can closely approximate target metrics.

\section{Conclusion}
\label{sec:conclusion}
In this study, we introduce Siesta, a novel framework designed to automatically synthesize proxy-apps for MPI programs. Siesta operates by tracing MPI programs to capture communication and computation events, symbolically representing these events, extracting behavioral patterns into a compact form, and translating this form into C code that replicates the original program's behaviors.
We assessed Siesta's effectiveness on a variety of MPI programs, including kernels, mini-apps, and production-level scientific computing software. 
The results demonstrate that Siesta accurately reflects the performance characteristics of MPI programs.

Siesta's generated proxy-apps still exhibit limitations compared to those manually developed by experts. Specifically, manually developed proxy-apps can adjust to varying input arguments and parallel scales, while Siesta's outputs are constrained to specific execution paths and fixed configurations. 
Synthetic proxy apps are more suitable for frequently updated or customized (not covered by common benchmarks) workloads, providing useful supplements to manually developed proxy apps.
The functionality of synthetic proxy-apps needs to be further enhanced in future work.



\section*{Acknowledgement}
This work is financially supported by National Natural Science Foundation of China (Grant number: 62202441) and Laoshan Laboratory (No.LSKJ202300305). The experiments in this paper were conducted on
the supercomputing systems in the Supercomputing Center of
University of Science and Technology of China and the National Supercomputing Center in Jinan.

\bibliography{main}
\bibliographystyle{IEEEtran}
\end{document}